\documentclass[a4paper]{jpconf}
\usepackage{graphicx}
\usepackage{amsmath}
\usepackage{amssymb}
\usepackage{physics}

\usepackage{amsmath,amsfonts,amsthm,bm}
\usepackage{braket}
\usepackage{subfig}

\begin{document}
\title{Fast-forward adiabatic quantum dynamics of XY spin model on three spin system}

\author{Iwan Setiawan$^{1}$, Riska Ekawita$^{2}$, Ryan Sugihakim$^{3}$, and Bobby Eka Gunara$^3$}
\address{$^1$ Department of Physics Education, University of Bengkulu, Kandang Limun, Bengkulu 38371, Indonesia}
\address{$^2$ Department of Physics, University of Bengkulu, Kandang Limun, Bengkulu 38371, Indonesia}
\address{$^3$ Department of Physics, Institut Teknologi Bandung, Jalan Ganesha 10, Bandung 40132, Indonesia}
\ead{iwansetiawan@unib.ac.id}
\date{\today}

\begin{abstract}
We discussed a method to accelerate an adiabatic quantum dynamics of $XY$ spin model by using the fast-forward method proposed by Masuda and Nakamura. The Accelerated scheme is constructed by adding the driving Hamiltonian to the original Hamiltonian and speeding it up with a large time-scaling factor and an adiabatic parameter that realizes adiabatic quantum dynamics in a shortened time. Accelerated adiabatic dynamics start by assuming the candidate of driving Hamiltonian consists of the pair-wise exchange interaction and magnetic field.  The driving Hamiltonian terms multiplied by the velocity function together with the original Hamiltonian give fast-forward driving for adiabatic states. We apply our method to $XY$ spin model by considering three spin systems on the Kagome lattice. In this model, we obtained the $XY$ pair-wise exchange interaction of nearest neighbors and next-nearest neighbors should be added to the original Hamiltonian as a driving interaction to accelerate the adiabatic motion. This pair-wise driving interaction in the fast-forward scheme guarantees the complete fidelity of accelerated states.
\end{abstract}

\section{Introduction}
The method to optimize and control the quantum evolution system has greatly improved quantum technology. Some techniques have been developed to obtain an optimum and robust quantum state. One of the techniques that attract much attention is adiabatic quantum evolution. In this technique, the eigenstate of the system would not change during a very small change of Hamiltonian, and it needs a longer time. Some methods have been developed to accelerate adiabatic quantum evolution. The method of shortcut to adiabaticity (STA) has received much more attention \cite{5,6,7,8,9,10,11,12,13,14}. The design of STA can be achieved using a variety of techniques\cite{10,14}. A universal approach is provided by counterdiabatic driving \cite{7}, also known as transitionless quantum driving. These methods have been developed to design a shortcut between two states without knowing the adiabatic relation between them \cite{14}. Instead of this method's work in quantum mechanics, related schemes appear in terms of classical mechanics \cite{jar} and stochastic dynamics \cite{ay}. Masuda and Nakamura \cite{1} constructed another method so-called fast-forward method. They consider to accelerate standard quantum dynamics by using the additional phase and driving potential to accelerate the wave function. The fast-forward method can speed up any given quantum system and obtain the target state on the shortened time scale by speeding up the reference quantum dynamics. Khujakulov and Nakamura \cite{1-1}  proposed that the fast-forward method is useful to improve the quantum tunneling power. This theory was further developed to speed up the quasi-static or adiabatic quantum dynamics \cite{2,3,4} by finding the regularization term and driving potential.  Some works on how to accelerate the adiabatic spin dynamics are also proposed \cite{15,16,17,18,19a} and extended to control the dynamics of many-body systems governed by the transverse Ising \cite{cam} and Lipkin-Meshkov-Glick \cite{ros} Hamiltonians. The fast-forward scheme of quantum entangled states \cite{19} is proposed by obtaining the driving interaction that consists of pair-wise interaction and magnetic field. The three body interactions for up to 4 spin systems \cite{20} should be added instead of pairwise interaction and magnetic field. The driving interactions in this paper depend on the geometry formed by the spin clusters. Application of the fast-forward method to three-spin of quantum annealing systems also have been reported \cite{19c}. This model is devoted to the triangle spin model and obtained the driving interaction that consists of pairwise interaction, magnetic field, and three-body interaction. The method to accelerate spin system on $XY$ model has been investigated\cite{15}. This paper considers an $XY$ spin model for two-spin and one-dimensional isotropic of many spin systems. By considering the spin interaction between the nearest neighbor of $X$ and $Y$ direction, they obtained the driving Hamiltonian to accelerate the dynamics of the system. Another method to accelerate the system using an $XY$ model in the regime in which the adiabatic theorem cannot be applied also have been reported\cite{15a}.

In this present work, we shall investigate the scheme of fast forward of adiabatic dynamics in $XY$ spin model by applying the scheme to three spin systems in a Kagome lattice. This spin system has a different time-dependent spin interaction between the nearest neighbors and the next-nearest neighbors. Using the fast-forward method, we obtain a  driving Hamiltonian to guarantee accelerated dynamics. In Section \ref{FFADspin}, we shall briefly summarize the scheme of fast forward of adiabatic quantum spin dynamics. In Section \ref{Kagome}, we consider the fast-forward Hamiltonian for $XY$ spin model for three spin systems in Kagome lattice. Section \ref{conc} is devoted to the conclusion.

\section{Fast-forward of adiabatic spin dynamics}\label{FFADspin}

In this section, we shall discuss a brief review of the theory of adiabatic spin dynamics proposed by Masuda and Nakamura \cite{19,19b}. Let us assume the adiabatic state of time-dependent  Schr\"{o}dinger equation (TDSE) as 
\begin{equation}\label{psi0}
\Psi_0(R(t)) =\begin{pmatrix}C_1(R) \\
\vdots \\
C_n(R)\end{pmatrix} e^{-\frac{i}{\hbar}\int_{0}^{t}E(R(t'))dt'} e^{i\xi(R(t))},
\end{equation}
where $C(R)$ is a spinor component, available from time independent Schr\"{o}dinger equation. Here 
\begin{equation}
R\equiv R(t)= R_0 + \epsilon t  
\end{equation}
is the adiabatically-changing parameter with $\epsilon \ll 1$, 
and $\xi$ is the adiabatic phase\cite{7}.
To obtain the adiabatic wave function as in Eq.(\ref{psi0}), we should  modify the Hamiltonian as regularized Hamiltonian as given by
\begin{equation}
H_0^{reg} = H_0 + \epsilon \mathcal{\tilde{H}},
\end{equation}
where $ \mathcal{\tilde{H}}$ is the regularization term.
The wave function as in Eq.(\ref{psi0}) and the regularized Hamiltonian will satisfy the time-dependent Schr\"{o}dinger equation 
\begin{equation}\label{schro6}
i \hbar \frac{\partial}{\partial_t}\Psi_0(R(t)) = H_0^{reg} \Psi_0(R(t)).
\end{equation}
By solving Eq.(\ref{schro6}) we obtain the equation to solve the regularization terms ($ \mathcal{\tilde{H}}$) in $O(\epsilon^1)$ as
\begin{eqnarray}\label{sum2}
\mathcal{\tilde{H}}_n\begin{pmatrix}C_1(R) \\
\vdots \\
C_n(R)\end{pmatrix}&=& i \hbar \partial_R\begin{pmatrix}C_1(R) \\
\vdots \\
C_n(R)\end{pmatrix}\nonumber\\
&& -\begin{pmatrix}C_1(R)^{\dagger} \\
\vdots \\
C_n(R)^{\dagger}\end{pmatrix} \partial_R \begin{pmatrix}C_1(R) \\
\vdots \\
C_n(R)\end{pmatrix}
\end{eqnarray}
which is the core equation of this scheme.
Here  $\mathcal{\tilde{H}}_n$ is the $n$-th state-dependent regularization term. The adiabatic wave function in Eq.(\ref{psi0}) will be obtained at the initial and the final time of evolution by using the regularization term as in Eq.(\ref{sum2}). Obtaining this adiabatic wave function needs a very long time, which will give less advantage to realize in the laboratory in particular when we want to see the time evolution of the wave function. To fast-forward such adiabatic motion, we introduce the time scaling factor ($\Lambda(t)$). The time scaling factor is a variable that can indicate an event occurring in a shorter time. By using the time scale factor, an event or motion of an object can be arranged according to the desired time. Several types of time scaling factors can be used for fast-forward schemes including spin cases such as cosine functions, hyperbolic functions, or Gaussian functions. This kind of function shortens the distance between two positions. In this three-spin system, the time scaling factor in the form of a cosine function is used. Here $\Lambda(t)$ which is defined as
\begin{equation}
\Lambda(t) = \int_{0}^{t} \alpha(t') dt'
\end{equation}
Here $\alpha$ is an arbitrary time magnification factor that is typically given by \cite{2}
\begin{equation}
\alpha(t) =\bar{\alpha}-(\bar{\alpha}-1)\cos \left(\frac{2 \pi}{T_{FF}}t\right)
\end{equation}
Time-dependent Schr\"{o}dinger equation for fast-forward state is written by
\begin{equation}
i \hbar \frac{\partial}{\partial_t} \Psi_{FF} = H_{FF} \Psi_{FF},
\end{equation}
where 
\begin{equation}\label{FF}
\Psi_{FF} = \begin{pmatrix}C_1(R(\Lambda(t))) \\
\vdots \\
C_n(R(\Lambda(t)))\end{pmatrix} e^{-\frac{i}{\hbar}\int_{0}^{t}E(R(\Lambda(t')))dt'} e^{i\xi(R(\Lambda(t)))}.
\end{equation}
By taking time derivative of $\Psi_{FF}$ in Eq.(\ref{FF}), we obtain the equation for fast-forward Hamiltonian as \cite{19}
\begin{eqnarray}\label{TDSE}
i \hbar \frac{\partial \Psi_{FF}}{\partial t} &=& \left(v(t) \mathcal{\tilde{H}}_n(R(\Lambda(t)))+ H_0(R(\Lambda(t)))
\right) \Psi_{FF}\nonumber\\
&\equiv&  H_{FF} \Psi_{FF},
\end{eqnarray}
where $v(t)$ is velocity function taken from the asymptotic limit,
limit $\bar{\alpha} \rightarrow \infty$ and $\epsilon \rightarrow 0$ under the constraint that $\bar{\alpha} \dot \epsilon \equiv \bar{v}$. Here $R(\Lambda(t))$ is given by \cite{19b}
\begin{equation}\label{R}
R(\Lambda(t))=R_0 + 2 \bar{v} \left(\frac{t}{2}-\frac{T_{FF} \sin \left(\frac{2 \pi  t}{T_{FF}}\right)}{4 \pi }\right).
\end{equation}
$H_{FF}$ is the driving Hamiltonian, $\mathcal{\tilde{H}}_n$ is the regularization term obtained from Eq.(\ref{sum2}) to generate the fast-forward scheme in spin system. As an example, we apply this fast-forward scheme to two spin systems with $XY$ spin model described in Appendix A.

\section{XY Spin model with Kagome Lattice}\label{Kagome}
\begin{figure}
\begin{center}
\includegraphics[width=1.5in]{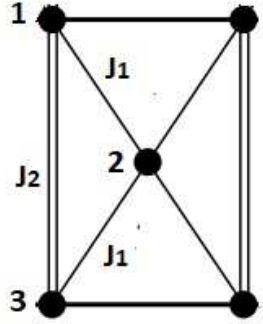}
\caption{Three spin systems of Kagome lattice}.\label{fig2}
\end{center}
\end{figure}

\begin{figure}
\begin{center}
  \includegraphics[width=2.5in]{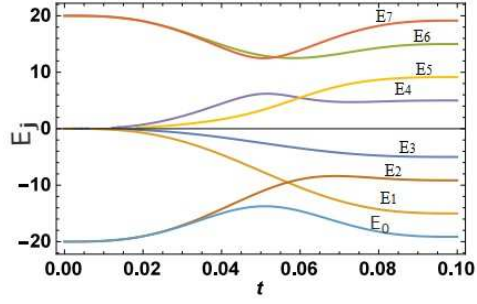}
\end{center}
\caption{Time dependence of  eigenvalue for $XY$ model in three spin system, with $J_1=J_0-R(\Lambda(t))$, $J_2=R(\Lambda(t))$, $B_z$=$B_0-R(\Lambda(t))$, $\bar{v}$ = 100, $T_{FF} = 0.1$, $J_0(R_0=0) = 10$, and $B_0 =0 $.} \label{fig3}
\end{figure}

To see how the above scheme works well, we also apply our method to three spin system models by considering the Kagome lattice   \cite{20,21,21a}. The configuration of this model is shown in Fig.(\ref{fig2}). Here, the time-dependent spin interactions between nearest neighbors, $J_1$ and next nearest neighbors, $J_2$ are shown by single and double bonds, respectively. This spin model has a different time-dependent spin interaction for nearest neighbor and next nearest neighbor which is described by a Kagome lattice, a two-dimensional lattice pattern found in the crystal structure of many natural minerals. Some minerals, namely jarosites and herbertsmithite, contain two-dimensional layers or three-dimensional Kagome lattice arrangements of atoms in their crystal structure. Here we consider a simple Kagome lattice for up to three spin systems. Furthermore, this method can be applied to accelerate the dynamics of microparticles on crystal structure that will be very useful for understanding and manipulating correlated quantum materials \cite{22}.
The Hamiltonian for three spin systems is given by
\begin{equation}\label{XY7}
H_0 = J_1 (\sigma_1^x \sigma_2^x+ \sigma_2^x \sigma_3^x)+J_2 (\sigma_3^y \sigma_1^y)+ \frac{1}{2}(\sigma_1^z+ \sigma_2^z+ \sigma_3^z)B_z(R(t)),
\end{equation}
with $J_1 = J_1 (R(t))$,  $J_2= J_2(R(t))$, and $B_z =B_z(R(t))$
By using these basis : $\Ket{\uparrow\uparrow\uparrow} $, $\Ket{\uparrow\uparrow\downarrow} $, $\Ket{\uparrow\downarrow\uparrow}$, $\Ket{\downarrow\uparrow\uparrow}$, $\Ket{\uparrow\downarrow\downarrow}$, $\Ket{\downarrow\uparrow\downarrow}$, $\Ket{\downarrow\downarrow\uparrow}$  and $\Ket{\downarrow\downarrow\downarrow}$. The matrix representation of Eq.(\ref{XY7}) is written as
\begin{equation}
H_0 = 
\begin{pmatrix}
 \frac{3 B_z}{2} & 0 & 0 & J_1 & 0 & -J_2 & J_1 & 0 \\
 0 & \frac{B_z}{2} & J_1 & 0 & J_2 & 0 & 0 & J_1 \\
 0 & J_1 & \frac{B_z}{2} & 0 & J_1 & 0 & 0 & -J_2 \\
 J_1 & 0 & 0 & -\frac{B_z}{2} & 0 & J_1 & J_2 & 0 \\
 0 & J_2 & J_1 & 0 & \frac{B_z}{2} & 0 & 0 & J_1 \\
 -J_2 & 0 & 0 & J_1 & 0 & -\frac{B_z}{2} & J_1 & 0 \\
 J_1 & 0 & 0 & J_2 & 0 & J_1 & -\frac{B_z}{2} & 0 \\
 0 & J_1 & -J_2 & 0 & J_1 & 0 & 0 & -\frac{3 B_z} {2} 
\end{pmatrix}.
\end{equation}
From the time-dependent eigenvalue of this Hamiltonian, as shown in Fig.(\ref{fig3}), we have numerically shown that there is no level crossing of the ground states $E_0$ of this model in the fast-forward time scale.
The ground state $E_0$ of this model is obtained as 
\begin{equation}
E_0 = -\frac{1}{2} \sqrt{3} \left(\frac{\beta }{\sqrt[3]{\Gamma }}-\sqrt[3]{\Gamma }\right)-\frac{\beta }{2 \sqrt[3]{\Gamma }}+\frac{B_z}{6}-\frac{\sqrt[3]{\Gamma }}{2}+\frac{J_2}{3},
\end{equation}
with \\ $\Gamma =\sqrt{\left(-\alpha -\frac{3 B_z^3}{16}-\frac{1}{27} \left(\frac{B_z}{2}+J_2\right)^3+\eta +\frac{J_2^3}{2}\right)^2-\beta ^3}+\gamma +\Delta$,\\
 $\alpha =\frac{1}{6} \left(\frac{B_z}{2}+J_2\right) \left(\frac{5 B_z^2}{4}-B_z J_2+4 J_1^2+J_2^2\right)$,\\
  $\beta =\frac{5 B_z^2}{12}-\frac{B_z J_2}{3}+\frac{1}{9} \left(\frac{B_z}{2}+J_2\right)^2+\frac{4 J_1^2}{3}+\frac{J_2^2}{3}$,\\
   $\gamma =-\frac{1}{8} 3 B_z^2 J_2-B_z J_1^2+\frac{B_z J_2^2}{4}-2 J_1^2 J_2$,\\
    $\eta =\frac{3 B_z^2 J_2}{8}+B_z J_1^2-\frac{B_z J_2^2}{4}+2 J_1^2 J_2$, and\\
     $\Delta =\alpha +\frac{3 B_z^3}{16}+\frac{1}{27} \left(\frac{B_z}{2}+J_2\right)^3-\frac{J_2^3}{2}$. 

The derivation of regularization term is given in Appendix C. From the matrix of the candidate of regularization term in Eq.(\ref{XY8}), the equation for regularization term is written as
\begin{eqnarray}\label{aku7}
&& i\hbar \frac{\partial  C_1}{\partial R}  =   \frac{3 \tilde{B}_z}{2} C_1 - 4i \tilde{W}_1 C_4 -2i \tilde{W}_2 C_6\nonumber\\
&& i\hbar \frac{\partial  C_4}{\partial R}  =  2 i \tilde{W}_1 C_1 -\frac{B_z}{2} C_4 ,\nonumber\\
&& i\hbar \frac{\partial  C_6}{\partial R}  =   2 i \tilde{W}_2 C_1 -\frac{\tilde{B}_z}{2} C_6,
\end{eqnarray}
which gives the solution
\begin{equation}
\tilde{B}_z= 0,
\end{equation}
\begin{equation}
\tilde{W_1}= -\frac{i \left(a C_4 C_1+3 b C_1^2-b C_6^2+c C_4 C_6\right)}{2 C_1 \left(3 C_1^2-2 C_4^2-C_6^2\right)},
\end{equation}
and
\begin{equation}
\tilde{W_2}= -\frac{i \left(a C_6 C_1+2 b C_4 C_6+3 c C_1^2-2 c C_4^2\right)}{2 C_1 \left(3 C_1^2-2 C_4^2-C_6^2\right)}.
\end{equation}
with $a= i\hbar \frac{\partial  C_1}{\partial R}$,  $b=i\hbar \frac{\partial  C_4}{\partial R}$, and $c= i\hbar \frac{\partial  C_6}{\partial R}$. Noting Eq.(\ref{adp2}), we find that $B_z = 0$. Comparing with two spin models, on this spin model we obtained that two regularization terms should be added to the original Hamiltonian i.e: $\tilde{W_1}$ and $\tilde{W_2}$
as an exchange parameter of $X$ and $Y$ direction of nearest neighbor and next nearest neighbor interaction respectively.

\begin{figure}
\subfloat[]{
  \includegraphics[width=2.5in]{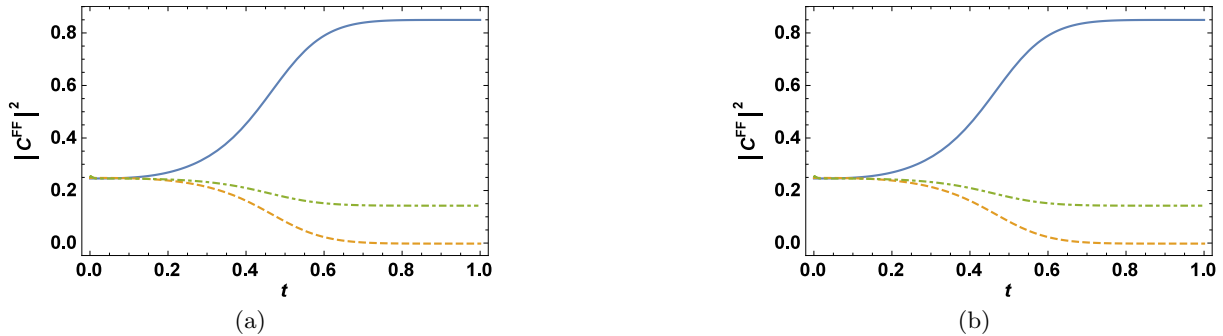}
}
\hfill
\subfloat[]{%
  \includegraphics[width=2.5in]{CFF.eps}%
}
\caption{Time dependence of  $|C_1^{FF}|^2$ (solid line), $|C_4^{FF}|^2$= $|C_7^{FF}|^2$(dashed line), and $|C_6^{FF}|^2$ (dot-dashed line: a. Obtaining by solving TDSE b. Obtaining from the eigenvector)} \label{fig4}
\end{figure}

\begin{figure}
\subfloat[]{%
  \includegraphics[width=2.5in]{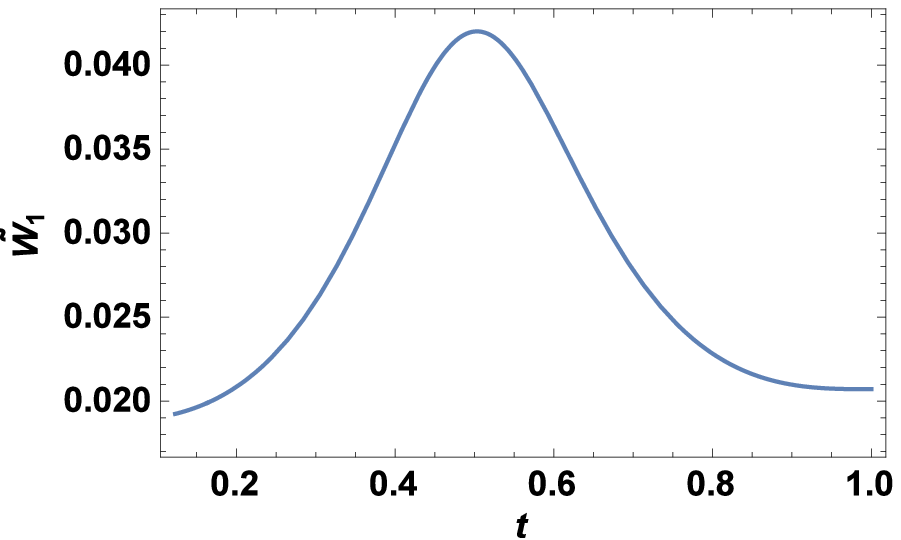}%
}
\hfill
\subfloat[]{%
  \includegraphics[width=2.5in]{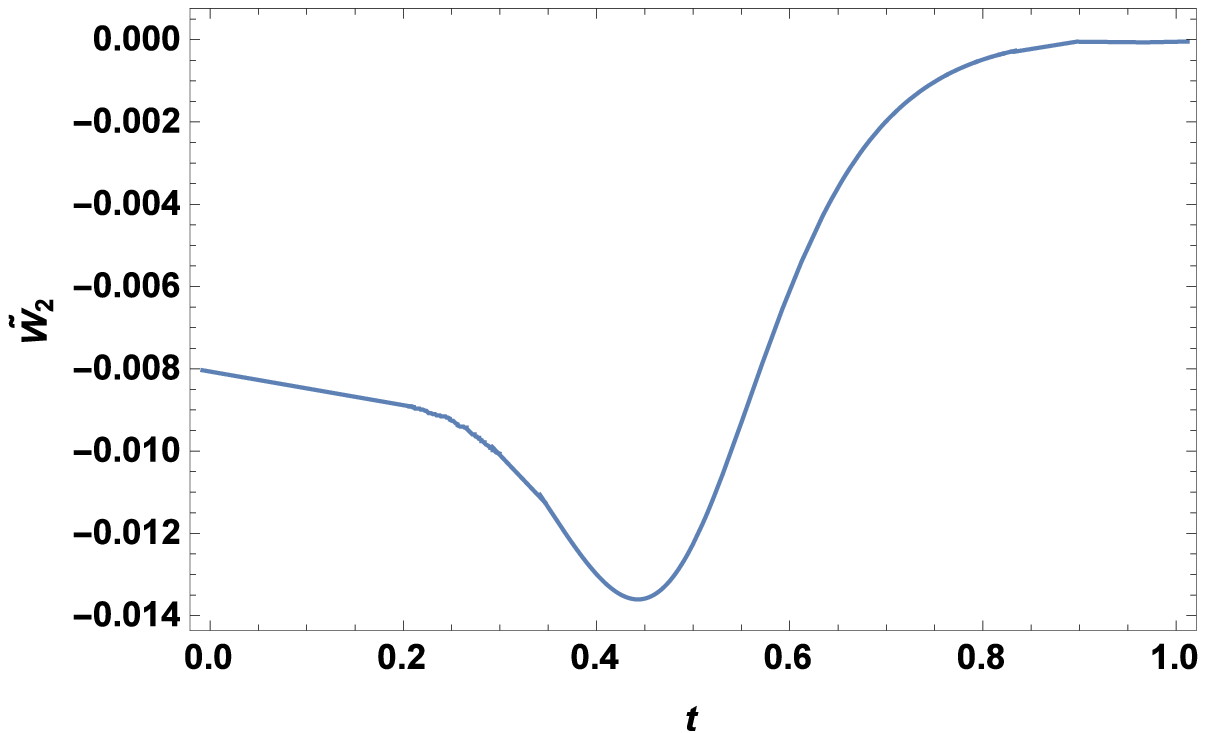}%
}
\caption{Time dependence of regularization term : (a) $\tilde{W}_1$; (b) $\tilde{W}_2$.} \label{fig5}
\end{figure}

By numerically solving the TDSE in Eq.(\ref{TDSE}) with parameter $R(\Lambda(t))$ in Eq.(\ref{R}), and $J_1=J_0-R(\Lambda(t))$, $J_2=R(\Lambda(t))$, $B_z$=$B_0-R(\Lambda(t))$, $\bar{v}$ = 10, $T_{FF} = 1$, $J_0(R_0=0) = 10$ and $B_0 =0 $, we see the dynamics of amplitude at the initial state as a linear combination of $\Ket{\uparrow\uparrow\uparrow} $,  $\Ket{\downarrow\uparrow\uparrow}$, $\Ket{\downarrow\uparrow\downarrow}$, and $\Ket{\downarrow\downarrow\uparrow}$. As $J_1$ and $B_x$ decreased from a positive value towards $0$ and $J_2$ increased from $0$ toward a positive value, the initial states rapidly change to nonentangled state i.e: $\Ket{\uparrow\uparrow\uparrow}$ state at the final time ($T_{FF}$) as shown in Fig.(\ref{fig4}). The solution of TDSE as shown in Fig.(\ref{fig4}) has a complete fidelity with time-dependent eigenvector as in Eq.(\ref{c1}, \ref{c2} and \ref{c3}),  during the fast forward time range $0\leq t\leq T_{FF} $. 
Time dependence of regularization term is shown in Fig.(\ref{fig5})

To conclude, for three spin systems in Kagome lattice, where the original Hamiltonian is an $XY$ interaction of the nearest neighbor with interaction parameter $J_1$ and $XY $ interaction of the next nearest neighbor with interaction parameter $J_2$, we obtain the regularization terms. These regularization terms consist of $\tilde{W}_1$ and $\tilde{W}_2$ as an exchange parameter of nearest neighbor (NN) interaction $(\sigma_1^x \sigma_2^y +\sigma_1^y \sigma_2^x)+(\sigma_2^x \sigma_3^y+ \sigma_2^y \sigma_3^x)$ and next nearest neighbor interaction (NNN)$(\sigma_1^x \sigma_3^y +\sigma_1^y \sigma_3^x)$ respectively. This regularization term should be added to the original Hamiltonian to guarantee the adiabatic motion and to speed up the dynamics of the spin.

\section{Conclusion}\label{conc}
We presented a scheme of fast-forward adiabatic spin dynamics proposed by Masuda and Nakamura to an $XY$ spin model in the Kagome lattice. We consider three spin systems with the original Hamiltonian, including time-dependent pair-wise exchange interaction parameters of nearest neighbor (NN), next nearest neighbor (NNN), and magnetic fields. We constructed the quasi-adiabatic dynamics by adding the regularization terms to the original Hamiltonian and then accelerated it using a large time-scaling factor. Assuming the candidate of regularization term that is experimentally realizable consisting of the exchange interactions and magnetic field, we solved the regularization terms. We obtained for this model, the regularization term are $\tilde{W}_1$ as an $xy$ exchange interaction parameter of $(\sigma_1^x \sigma_2^y +\sigma_1^y \sigma_2^x)+(\sigma_2^x \sigma_3^y+ \sigma_2^y \sigma_3^x)$ and $\tilde{W}_2$  for $(\sigma_1^x \sigma_3^y +\sigma_1^y \sigma_3^x)$. The candidate of regularization term for three spin system of an $XY$  model in Kagome lattice can be obtained from the candidate of regularization term in two spin system by adding an $XY$ pair-wise exchange interaction in the nearest neighbor (NN) and next-nearest neighbor (NNN)  as $\mathcal{\tilde{H}} =  \sum_{(i,j)\in all } \tilde{W}_{ij}^{NN} (\sigma_i^x \sigma_j^y +\sigma_i^y \sigma_j^x) +\sum_{(i,j)\in all } \tilde{W}_{ij}^{NNN} (\sigma_i^x \sigma_j^y +\sigma_i^y \sigma_j^x) $. We also confirmed the complete fidelity of the wave function during the fas-forward motion. Furthermore, this method will be beneficial to accelerate the dynamics of microparticles on crystal structures that form a Kagome geometry. A further investigation of many $XY$ spin systems of Kagome lattice remains a challenge.

\ack
I.S is grateful to Prof. Katsuhiro Nakamura, Rudy Kusdiantara, Ardian Nata Atmaja, Andy Octavian Latif, and Fiki Taufik for a valuable discussion. I.S is supported by PDUPT grant from Kemendikburistek 2022. B.E.G acknowledged Hibah PPD Kemdikbud Ristek for financial support.

\appendix
\renewcommand{\theequation}{A.\arabic{equation}}
\section*{Appendix A. An $XY$ spin model of two spin systems}\label{ap}
This model is a Heisenberg spin interaction with $X$ and $Y$ directions and an external magnetic field in $Z$ direction. Furthermore, the model can be applied to two-qubit systems that are very important in developing a quantum computer. The Hamiltonian for this model is written as \cite{15}
\begin{equation}\label{XY1}
H_0 = J_1(R(t)) \sigma_1^x \sigma_2^x + J_2(R(t)) \sigma_1^y \sigma_2^y+ \frac{1}{2}(\sigma_1^z+ \sigma_2^z)B_z(R(t)),
\end{equation}
On these two spin systems, the spin interacts with the nearest neighbor where $J_1$ and $J_2$ are the time-dependent interaction parameters, and $\sigma^x$ and $\sigma^y$ are Pauli matrices. By using the basis on two spin systems : $\Ket{\uparrow\uparrow} $, $\Ket{\uparrow\downarrow} $, $\Ket{\downarrow\uparrow}$, and $\Ket{\downarrow\downarrow}$, matrix representation of Eq.(\ref{XY1}) can be written as
\begin{equation}\label{XY2}
H_0 = \left(
\begin{array}{cccc}
B_z & 0 & 0 & J_1-J_2 \\
0  & 0 & J_1+J_2 & 0 \\
 0 & J_1+J_2 & 0 & 0 \\
 J_1-J_2 & 0 & 0 & -B_z \\
\end{array}
\right).
\end{equation}
We can see from the Hamiltonian in Eq. (\ref{XY2}), time-dependent of eigenvalue for this Hamiltonian as shown in Fig.(\ref{fig6})
\renewcommand{\thefigure}{A\arabic{figure}}

\begin{figure}
\begin{center}
  \includegraphics[width=2.5in]{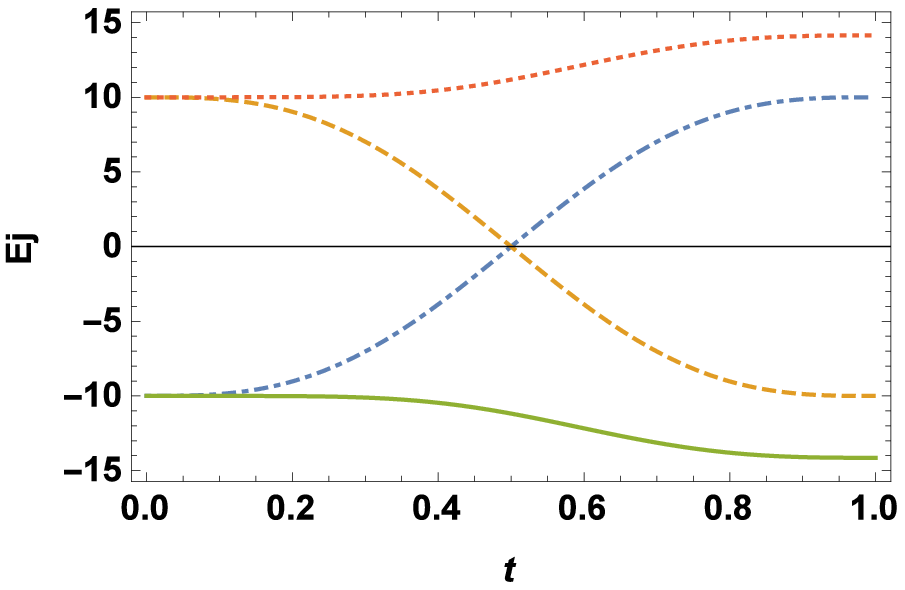}
\end{center}
\caption{Time dependence of  eigenvalue (solid line for the ground state) for $XY$ model in two spin systems, with $J_1=J_0-R(\Lambda(t))$, $J_2=R(\Lambda(t))$, $B_z$=$B_0-R(\Lambda(t))$, $\bar{v}$ = 10, $T_{FF} = 1$, $J_0(R_0=0) = 10$, and $B_0 =0 $.} \label{fig6}
\end{figure}
\begin{figure}
\subfloat[]{%
  \includegraphics[width=2.5in]{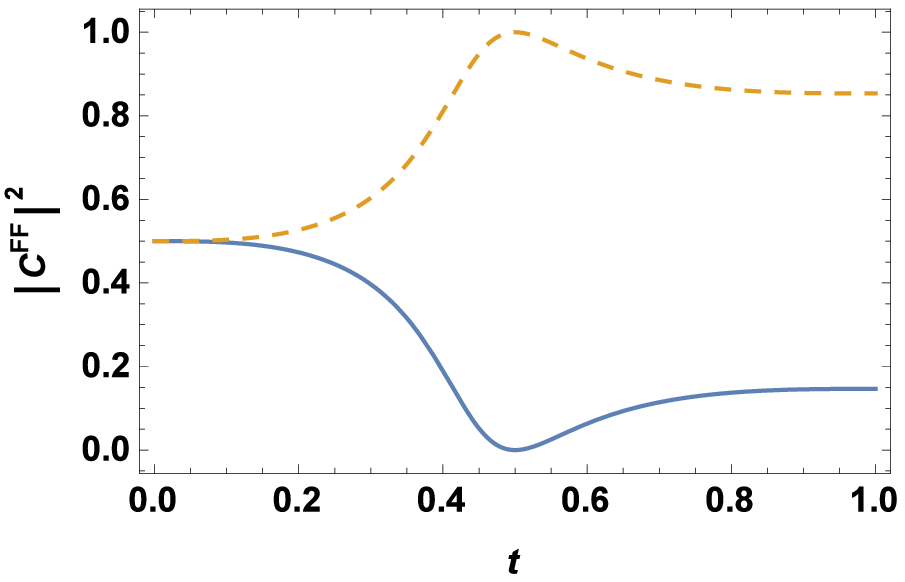}%
}
\hfill
\subfloat[]{%
  \includegraphics[width=2.5in]{Schro4.eps}%
}
\caption{Time dependence of $|C_1^{FF}|^2$ (solid line), $|C_4^{FF}|^2$ (dotted line) with the same parameter as in Fig.(\ref{fig6}) :(a) obtained by Solving TDSE in Eq.{\ref{TDSE}}  (b) obtained from the eigenvector} \label{fig1}
\end{figure}
\begin{figure}
\begin{center}
  \includegraphics[width=2.5in]{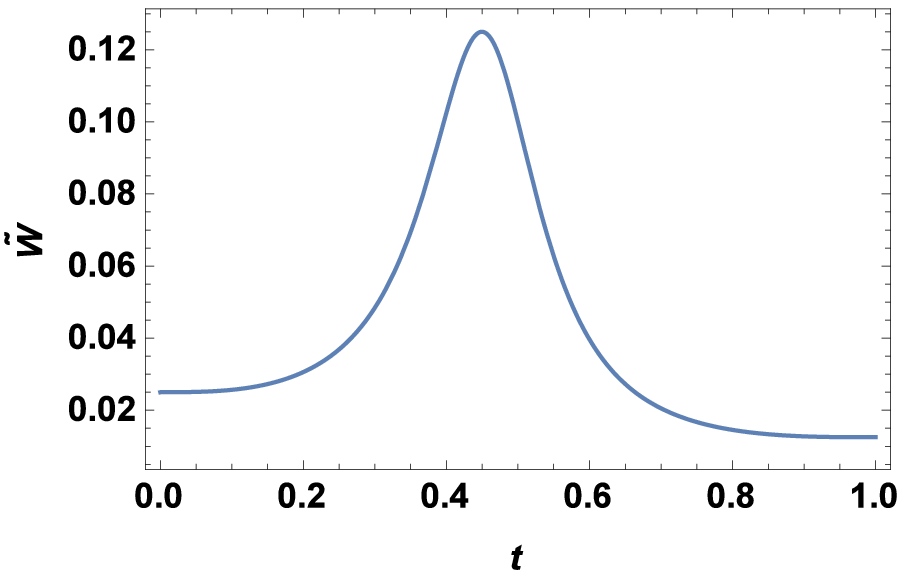}
\end{center}
\caption{Time dependence of  of regularization term, with $J_1=J_0-R(\Lambda(t))$, $J_2=R(\Lambda(t))$, $B_z$=$B_0-R(\Lambda(t))$, $\bar{v}$ = 10, $T_{FF} = 1$, $J_0(R_0=0) = 10$, and $B_0 =0 $.} \label{fig1b}
\end{figure}

By decreasing $J_1$ and $B_z$ from a positive value towards $0$ and increasing $J_2$ from $0$ towards a positive value, the ground state of the system as indicated by a solid line in Fig.(\ref{fig6}), will have no transition from the initial time and the final time of the evolution. 
Here we consider applying the fast-forward scheme to the ground state of Hamiltonian in Eq.(\ref{XY2}). This ground state is obtained as $E_0 = -\sqrt{B_z^2+J_1^2+J_2^2-2 J_1 J_2}$. Here we shall obtain the driving Hamiltonian for these two spin systems by using the equation of regularization term in Eq.(\ref{sum2}). To solve the regularization term, we use a candidate of regularization term as written by
\begin{eqnarray}\label{tH1}
\mathcal{\tilde{H}} = \tilde{W}(\sigma_1^x \sigma_2^y +\sigma_1^y \sigma_2^x)+\frac{1}{2}(\sigma_1^z+\sigma_2^z)\tilde{B}_z,
\end{eqnarray}  
The detailed derivation to obtain the regularization term is given in Appendix C. We obtained the solution for regularization term equation (as written in Eq.(\ref{XY6})) as 
\begin{equation}
\tilde{B}_z = 0,
\end{equation}
and
\begin{equation}
\tilde{W}=-\frac{i \left(a C_4+b C_1\right)}{2 \left(C_1^2-C_4^2\right)},
\end{equation}
with $a =i \hbar \frac{\partial C_1}{\partial_R} $ and $b = i \hbar \frac{\partial C_4}{\partial_R}$. Here by noting Eq.(\ref{adp}) we find $B_z = 0$.  The matrix expression for regularization term is given by
\begin{equation}
\mathcal{\tilde{H}} = \begin{pmatrix} 
0 & 0 & 0 & -i \tilde{W} \\
0 & 0 & 0 & 0 \\
0 & 0 & 0 & 0 \\
i \tilde{W} & 0 & 0 & 0
 \end{pmatrix},
\end{equation}
with $\tilde{W}$ is the exchange interaction parameter for $\sigma_1^x\sigma_2^y+\sigma_1^y\sigma_2^x$. This regularization term is analogous to the result obtained by Takahashi \cite{15}. By using an explicit expression of eigenvector ($C_1$ and $C_2$) in Eq.(\ref{eig1}) and Eq.(\ref{eig2}), together with its $R$ derivative, the regularization term $\tilde{W}$ can be rewritten as
\begin{equation}\label{reg}
\tilde{W} =\frac{B_z (\dot{J_1}-\dot{J_2})+\dot{B_z} (J_2-J_1)}{2 \left(B_z^2+(J_1-J_2)^2\right)}.
\end{equation}
Using Eq.(\ref{TDSE}), the fast forward Hamiltonian is given by
\begin{eqnarray}
H_{FF} &=& J_1(\Lambda(t)) \sigma_1^x \sigma_2^x + J_2(\Lambda(t)) \sigma_1^y \sigma_2^y\nonumber\\
&+& \frac{1}{2}(\sigma_1^z+ \sigma_2^z)B_z(\Lambda(t))
+ v(t)\tilde{W}.
\end{eqnarray}
We then numerically solved the fast forward Schr\"{o}dinger equation in Eq.(\ref{TDSE}) with the additional regularization term as in Eq.(\ref{reg}) instead of the original Hamiltonian in Eq.(\ref{XY1})  with parameter $R(\Lambda(t))$ in Eq.(\ref{R}). Here we used parameter $J_1=J_0-R(\Lambda(t))$, $J_2=R(\Lambda(t))$, $B_z$=$B_0-R(\Lambda(t))$, $\bar{v}$ = 10, $T_{FF} = 1$, $J_0(R_0=0) = 10$ and $B_0 =0 $. We see by adding the regularization term to the original Hamiltonian and decreasing $J_1$ and $B_z$ from a positive value towards $0$ and increasing $J_2$ from $0$ towards a positive value, the dynamics of the state at the initial time as a linear combination of $\Ket{\uparrow\uparrow}$ and $\Ket{\downarrow\downarrow}$ states change rapidly to nonentangled state at the final time ($T_{FF}$). The dynamics of wave function $\Psi_{FF}$ of fast forward Schr\"{o}dinger equation is exactly the same as the dynamics of wave function in the standard Schr\"{o}dinger equation (without fast forwarding), i.e the eigenstate of the Hamiltonian as in Eq.(\ref{eig1}) and Eq.(\ref{eig2}). This fact is shown in Fig.({\ref{fig1}}). This result showed that by adding the regularization term, we can get the desired adiabatic wave function in a shorter time. The time dependence of regularization term ($\tilde{W}$) is shown in Fig.(\ref{fig1b})

We confirmed that we obtained the driving Hamiltonian for two spin systems of $XY$ model. By adding the regularization term $\tilde{W}$ to the original Hamiltonian, we obtain the complete fidelity. This guarantee that the regularization term $\tilde{W}$ can accelerate the system adiabatically.

\section*{Appendix B. The derivation of regularization term for two spin systems}\label{App}

The normalized component of eigenvector for the ground state is written as
\begin{equation}\label{eig1}
C_1 = \kappa \Bigg(\sqrt{\frac{J_1^2-2J_1J_2+J_2^2}{2B_z^2+2J_1^2+2J_2^2-4J_1J_2-2B_z\omega}}\Bigg),
\end{equation}
\begin{equation}
C_2 = C_3 = 0,
\end{equation}
and
\begin{equation}\label{eig2}
C_4 = \sqrt{\frac{J_1^2-2J_1J_2+J_2^2}{2B_z^2+2J_1^2+2J_2^2-4J_1J_2-2B_z\omega}}, 
\end{equation}
with 
\begin{equation}
\kappa =   -\frac{\sqrt{B_z^2+J_1^2+J_2^2-2 J_1 J_2}-B_z}{J_1-J_2},
\end{equation} 
and
\begin{equation}
\omega = \sqrt{B_z^2+J_1^2-2J_1J_2+J_2^2}.
\end{equation}
By using Eq.(\ref{sum2}) and the component of eigenvector, the equation to obtain regularization term can be rewritten as
\begin{eqnarray}\label{xy3}
&& i\hbar \frac{\partial  C_1}{\partial R}  =   \tilde{\mathcal{H}}_{11}C_1 + \tilde{\mathcal{H}}_{14} C_4 \\\nonumber
&&
i\hbar \frac{\partial  C_2}{\partial R} =                \tilde{\mathcal{H}}_{21}C_1+\tilde{\mathcal{H}}_{24} C_4=0     \\\nonumber
&& i\hbar \frac{\partial  C_3}{\partial R} =  \tilde{\mathcal{H}}_{31} C_1+ \tilde{\mathcal{H}}_{34} C_4 =0 \\\nonumber  
&&
 i\hbar \frac{\partial  C_4}{\partial R}  =    \tilde{\mathcal{H}_{41}}C_1 + \tilde{\mathcal{H}}_{44}C_4.
\end{eqnarray}
From the normalization $C_1^2+C_4^2 = 1 $, we see that
\begin{equation}\label{adp}
C_1 \frac{\partial C_1}{\partial_R}+C_4 \frac{\partial C_4}{\partial_R}=0,
\end{equation} 
and then adiabatic phase $\xi $= 0.
To solve the regularization term in Eq.(\ref{xy3}), we use a candidate of regularization term as written in Eq. (\ref{tH1}), with $\tilde{W}$ is pair-wise exchange parameter of $(\sigma_1^x \sigma_2^y +\sigma_1^y \sigma_2^x)$.
The matrix representation of Eq.(\ref{tH1}) is obtained as
\begin{equation}\label{XY5}
\mathcal{\tilde{H}} = \begin{pmatrix} 
  \tilde{B}_z & 0 & 0 & -i2 \tilde{W_1}  \\
  0 &  0 & 0 & 0  \\
 0 & 0 & 0 & 0 \\
  i2\tilde{W}_1 & 0 & 0 & -\tilde{B}_z
 \end{pmatrix}.
\end{equation}
By substituting Eq.(\ref{XY5}) to Eq.(\ref{xy3}), we obtain
\begin{eqnarray}\label{XY6}
&& i\hbar \frac{\partial  C_1}{\partial R}  =   \tilde {B}_z C_1 - 2 i \tilde{W} C_4 \\\nonumber
&&
 i\hbar \frac{\partial  C_4}{\partial R}  =    2 i \tilde{W} C_1 - \tilde{B}_z C_4,
\end{eqnarray}
\renewcommand{\theequation}{B.\arabic{equation}}
\section*{Appendix C. The derivation of regularization term for three spin systems}\label{App1}

From the eigen vector of the ground state, we see that $C_1 \neq 0$, $C_2=C_3=C_5=C_8=0$, $C_4=C_7 \neq 0$, and $C_6 \neq 0$. The normalized component of eigenvector for the ground state is written as
\begin{equation}\label{c1}
C_1 = B_1 \zeta,
\end{equation}

\begin{equation}\label{c2}
C_4 = C_7= B_4 \zeta,
\end{equation}

\begin{equation}\label{c3}
C_6 = B_6 \zeta,
\end{equation}
where, 
\begin{eqnarray}
B_1 &=& 
-\frac{-B_z^2+16 J_1^2+4 J_2^2}{8 B_z J_1}\nonumber\\
&+&\frac{\left(\frac{1}{2} \sqrt{3} i \left(\frac{\beta }{\sqrt[3]{\Gamma }}-\sqrt[3]{\Gamma }\right)+\frac{\beta }{2 \sqrt[3]{\Gamma }}-\frac{B_z}{6}+\frac{\sqrt[3]{\Gamma }}{2}-\frac{J_2}{3}\right)^2}{2 B_z J_1}\nonumber\\
&-&\frac{\frac{1}{2} \sqrt{3} i \left(\frac{\beta }{\sqrt[3]{\Gamma }}-\sqrt[3]{\Gamma }\right)+\frac{\beta }{2 \sqrt[3]{\Gamma }}-\frac{B_z}{6}+\frac{\sqrt[3]{\Gamma }}{2}-\frac{J_2}{3}}{2 J_1},\nonumber\\
\end{eqnarray}
\begin{equation}
B_4 = 1,
\end{equation}
\begin{eqnarray}
B_6 &=& \frac{3 B_z^2-8 B_z J_2+16 J_1^2+4 J_2^2}{8 B_z J_1}\nonumber\\
&-&\frac{\left(\frac{1}{2} \sqrt{3} i \left(\frac{\beta }{\sqrt[3]{\Gamma }}-\sqrt[3]{\Gamma }\right)+\frac{\beta }{2 \sqrt[3]{\Gamma }}-\frac{B_z}{6}+\frac{\sqrt[3]{\Gamma }}{2}-\frac{J_2}{3}\right)^2}{2 B_z J_1}\nonumber\\
&-&\frac{\frac{1}{2} \sqrt{3} i \left(\frac{\beta }{\sqrt[3]{\Gamma }}-\sqrt[3]{\Gamma }\right)+\frac{\beta }{2 \sqrt[3]{\Gamma }}-\frac{B_z}{6}+\frac{\sqrt[3]{\Gamma }}{2}-\frac{J_2}{3}}{2 J_1},\nonumber\\
\end{eqnarray}
and
\begin{equation}
\zeta = \frac{1}{\sqrt{B_1^2+B_4^2+B_6^2+B_7^2}}.
\end{equation}
By using Eq.(\ref{sum2}) and component of eigenvector, the equation for regularization term can be rewritten as
\begin{eqnarray}\label{aku6}
&& i\hbar \frac{\partial  C_1}{\partial R}  =   \tilde{\mathcal{H}}_{11}C_1 + (\tilde{\mathcal{H}}_{14} +\tilde{\mathcal{H}}_{17}) C_4 +\tilde{\mathcal{H}}_{16}C_6\nonumber\\
&& i\hbar \frac{\partial  C_2}{\partial R}  =   \tilde{\mathcal{H}}_{21}C_1 + (\tilde{\mathcal{H}}_{24} +\tilde{\mathcal{H}}_{27}) C_4 +\tilde{\mathcal{H}}_{26}C_6=0 ,\nonumber\\
&& i\hbar \frac{\partial  C_3}{\partial R}  =   \tilde{\mathcal{H}}_{31}C_1 + (\tilde{\mathcal{H}}_{34} +\tilde{\mathcal{H}}_{37}) C_4 +\tilde{\mathcal{H}}_{36}C_6=0,\nonumber\\
&& i\hbar \frac{\partial  C_4}{\partial R}  =   \tilde{\mathcal{H}}_{41}C_1 + (\tilde{\mathcal{H}}_{44} +\tilde{\mathcal{H}}_{47}) C_4 +\tilde{\mathcal{H}}_{46}C_6,\nonumber\\
&& i\hbar \frac{\partial  C_5}{\partial R}  =   \tilde{\mathcal{H}}_{51}C_1 + (\tilde{\mathcal{H}}_{54} +\tilde{\mathcal{H}}_{57}) C_4 +\tilde{\mathcal{H}}_{56}C_6=0 ,\nonumber\\
&& i\hbar \frac{\partial  C_6}{\partial R}  =   \tilde{\mathcal{H}}_{61}C_1 + (\tilde{\mathcal{H}}_{64} +\tilde{\mathcal{H}}_{67}) C_4 +\tilde{\mathcal{H}}_{66}C_6 ,\nonumber\\
&& i\hbar \frac{\partial  C_7}{\partial R}  =   \tilde{\mathcal{H}}_{71}C_1 + (\tilde{\mathcal{H}}_{74} +\tilde{\mathcal{H}}_{77}) C_4 +\tilde{\mathcal{H}}_{76}C_6 ,\nonumber\\
&& i\hbar \frac{\partial  C_8}{\partial R}  =   \tilde{\mathcal{H}}_{81}C_1 + (\tilde{\mathcal{H}}_{84} +\tilde{\mathcal{H}}_{87}) C_4 +\tilde{\mathcal{H}}_{86}C_6=0 .\nonumber\\.
\end{eqnarray}
From the normalization factor 
\begin{equation}
C_1^2 + 2 C_4^2+C_6^2 = 1
\end{equation}
we see that
\begin{equation}\label{adp2}
C_1  \frac{\partial  C_1}{\partial R}+2 C_4  \frac{\partial  C_4}{\partial R}+C_6  \frac{\partial  C_6}{\partial R}=0.
\end{equation}
To solve this equation, a candidate of regularization as shown in Eq.(\ref{tH1})  is also proposed, i.e :
\begin{eqnarray}\label{cdcand}
\mathcal{\tilde{H}}& =& \tilde{W}_1 [(\sigma_1^x \sigma_2^y +\sigma_1^y \sigma_2^x)+(\sigma_2^x \sigma_3^y+ \sigma_2^y \sigma_3^x)]\nonumber\\
&+&\tilde{W}_2 (\sigma_3^x \sigma_1^y+ \sigma_3^y \sigma_1^x)
+ \frac{1}{2}(\sigma_1^z+\sigma_2^z+\sigma_3^z)\tilde{B}_z,
\end{eqnarray}
where $\tilde{W}_1$ is a time dependent $XY$ interaction parameter of nearest neighbor  and $\tilde{W}_2$ is for next nearest neighbor
The matrix representation of the candidate of regularization term in Eq.(\ref{cdcand}) is given by

\begin{eqnarray}\label{XY8}
\mathcal{\tilde{H}} =
\begin{pmatrix}
 \frac{3 \tilde{B}_z}{2} & 0 & 0 & -2 i \tilde{W}_1 & 0 & -2 i \tilde{W}_2 & -2 i \tilde{W}_1 & 0 \\
 0 & \frac{\tilde{B}_z}{2} & 0 & 0 & 0 & 0 & 0 & -2 i \tilde{W}_1 \\
 0 & 0 & \frac{\tilde{B}_z}{2} & 0 & 0 & 0 & 0 & -2 i \tilde{W}_2 \\
 2 i \tilde{W}_1 & 0 & 0 & -\frac{\tilde{B}_z}{2} & 0 & 0 & 0 & 0 \\
 0 & 0 & 0 & 0 & \frac{B_z}{2} & 0 & 0 & -2 i \tilde{W}_1 \\
 2 i \tilde{W}_2 & 0 & 0 & 0 & 0 & -\frac{\tilde{B}_z}{2} & 0 & 0 \\
 2 i \tilde{W}_1 & 0 & 0 & 0 & 0 & 0 & -\frac{\tilde{B}_z}{2} & 0 \\
 0 & 2 i \tilde{W}_1 & 2 i \tilde{W}_2 & 0 & 2 i \tilde{W}_1 & 0 & 0 & -\frac{ 3 \tilde{B}_z}{2}
 \end{pmatrix}. \nonumber\\
\end{eqnarray}

\end{document}